\documentstyle[multicol,aps,prl]{revtex}
\begin{document}
\draft
\preprint{}
\title{Analytical Results for Nontrivial Polydispersity Exponents in 
Aggregation Models}

\author{St\'ephane Cueille and Cl\'ement Sire}
\address{Laboratoire de Physique Quantique (UMR C5626 du CNRS),
Universit\'e Paul Sabatier\\
31062 Toulouse Cedex, France.\\}
\maketitle
\begin{abstract}
We study a Smoluchowski equation describing a simple mean-field model of
particles moving in $d$ dimensions and aggregating with conservation of `mass'
$s=R^D$ ($R$ is the particle radius). In the scaling regime the scaled mass
distribution $P(s)\sim s^{-\tau}$, and $\tau$ can be computed by perturbative
and non perturbative expansions. A possible application to two-dimensional
decaying turbulence is briefly discussed.

\end{abstract}



\begin{multicols}{2}

Aggregation phenomena are widespread in nature. They have such an impact on
material sciences, chemistry, astrophysics, that a large amount of literature
was devoted to them \cite{meakinrev}.  In such dynamical processes, particles
or objects as  different in geometry and size as colloidal particles,
galaxies, small molecules, vortices, droplets, polymers, can merge to form a
new entity when they come into close contact or interpenetrate, through
diffusion (Brownian coagulation \cite{smolu,dlcca}), ballistic motion
(ballistic agglomeration \cite{pomeau,KRL,trizac}), exogenous growth (droplets
growth and coalescence \cite{meakindrop}) or droplet deposition
\cite{familymeakin}.

We are usually interested in the evolution of the statistical distribution of
the `mass' $s$, a quantity characteristic of each particle, that is conserved
in the  coalescence process: it can be either the real mass, the volume, the
area, the electric charge, or any other parameter depending on the underlying
physics.

Great advance was achieved when it was proposed \cite{vicsek} and observed
both in real experiments and in numerical simulations that the distribution
$N(s,t)$ exhibits scale invariance at large time: 
\begin{equation}
N(s,t)\sim S(t)^{-\beta} f\left(\frac{s}{S(t)}\right) 
\end{equation} 
where the characteristic mass $S(t)$ diverges as $t^z$  when $t\to\infty$,
ensuring the oblivion of initial  conditions and physical cut-off or
discreteness, as does the diverging correlation length of critical phenomena:
universality arises in dynamics as well, with new universality classes.

The exponents $z$ and $\beta$ are easily deduced from conservation laws and
physical arguments, but in many cases a polydispersity exponent $\tau$ defined
by $f(x)\sim x^{-\tau}$ when $x\rightarrow 0$ is observed, whose value is
nontrivial though universal. The prediction of $\tau$ is still a challenge.

The earliest tool for tackling the problem is still one of the most popular,
that is the Smoluchowski equation \cite{smolu}. It is a master equation
\cite{vankam} for the distribution $N(s,t)$:
\begin{eqnarray} \label{smoleq}
\frac{\partial N(s,t)}{\partial t}
&=\frac{1}{2}\int N(s_1,t)N(s-s_1,t)K(s_1,s-s_1)\,ds_1\nonumber\\
&- N(s,t) \int N(s_1,t) K(s,s_1) \,ds_1
\end{eqnarray} 
where the aggregation kernel $K(x,y)$ is symmetric and is characteristic of
the physics of the aggregation process on a more or less coarse-grained level.
Such kinetic equations are usually derived within a mean-field approximation,
but in certain cases it is possible to go beyond mean-field limitations
investigated by van Dongen \cite{vandong1} by including approximately  spatial
correlation effects in the kernel $K$ \cite{poland,nous}.

One would like to be able to extract the nontrivial exponent $\tau$  for the
specific system from the proper kinetic equation. This is not an easy task
however: the problem was clarified by van Dongen and Ernst \cite{vandong2} who
classified the kernels according to their homogeneity and asymptotic behavior:
\begin{eqnarray}\label{genscal1}
K(bx,by)=b^\lambda K(x,y)\\
K(x,y)\sim x^\mu y^\nu \,\,\, (y \gg x) \label{genscal2}
\end{eqnarray}
For a given physical system, the homogeneity  $\lambda$ is easily determined
using scaling arguments. We consider only nongelling  systems with
$\lambda\leq1$ \cite{vandong2}. For $\mu>0$, the exponent $\tau$ is trivial
and found to  be $\tau=1+\lambda$, whereas for $\mu=0$, $\tau$ depends on the
whole solution $f$ of the scaling equation derived from Eq. (\ref{smoleq})
(see Eq. (\ref{scaleq}) below). $\mu<0$ does not lead to any  power law
behavior but rather to a bell-shaped scaling function $f$ \cite{vandong2}. In
the following, we shall focus on the $\mu=0$ case  for which  the exponent
$\tau$ has been so far only determined  numerically by direct simulation of
Smoluchowski equation (not an easy task) \cite{krivitsky}, by  time series
\cite{song}, and of course by direct simulation of the physical system
described by the considered Smoluchowski equation
\cite{meakinrev,dlcca,pomeau,trizac,meakindrop,familymeakin,vicsek}.
Considering the ubiquity and the importance of the $\mu=0$ case leading to
nontrivial polydispersity exponents,  analytical results would be certainly
welcome.   

The purpose of this Letter is to show, working with a physically relevant
simple kernel, that some information about  $\tau$ can be extracted from the
kernel itself using exact bounds, estimates  and expansions around exactly
solvable kernels. We compare our results to numerical studies in the
literature \cite{krivitsky,song} and briefly discuss a possible original
application to two-dimensional decaying turbulence.

Consider hyperspherical particles in a $d$-dimensional box, of polydisperse
radii $R$ with distribution $F(R,t)$, evolving the following way: at time $t$
we choose the positions of their centers with uniform probability in
$d$-space. Then each pair of overlapping spheres of radii $R_1$ and $R_2$
merges to form a new sphere of radius,
\begin{equation}\label{conser}
R=(R_1^{D}+R_2^{D})^\frac{1}{D}
\end{equation} 
where $D$ is a parameter with $D\geq d$. $D$ can be the actual dimension of
the spheres, as for instance in the case of $D=3$ spheres deposited on a $d=2$
plane \cite{familymeakin}. Once each coalescence has been resolved, we have
reached $t+1$.

The conserved variable is $s=R^D$ and the corresponding kernel for the
equation (\ref{smoleq}) is $K(x,y)=(x^\frac{1}{D}+y^\frac{1}{D})^d$. This
kernel has been introduced in many contexts from molecular coagulation
\cite{poland} to cosmology \cite{krivitsky,silk} for specific values of $d$
and $D$, and is one of the most studied in the literature
\cite{poland,vandong2,krivitsky,song,silk,ziff,vandong3,Leyvraz} although very
few analytical results are known. This kernel has $\lambda=\frac{d}{D}$ and
$\mu=0$.  From the conservation law we get \cite{vandong2} $\beta=2$ and if we
plug the scaling form into (\ref{smoleq}) we get $z=D/(D-d)$ for $d<D$  and: 
\begin{eqnarray}\label{scaleq}
&sf'(s)+2f(s)=f(s)\int_0^{+\infty} f(s_1)(s_1^\frac{1}{D}
+s^\frac{1}{D})^d ds_1  \nonumber\\ 
&- \frac{1}{2}\int_0^s f(s_1)f(s-s_1)(s_1^\frac{1}{D}
+(s-s_1)^\frac{1}{D})^d \,ds_1
\end{eqnarray}
If $\tau\geq 1$ each term of the RHS of Eq. (\ref{scaleq}) is separately
divergent and they should be properly grouped \cite{nous,vandong2}.

In $d=0$ or $D=\infty$, Eq. (\ref{scaleq}) reduces to the constant kernel
equation with exact solution $f_0(x)=2e^{-s}$ and
$f_{\infty}(s)=2^{1-d}e^{-s}$. In the case $d=1$, $D=1$ an exact analytic
solution is also known for the time dependent equation, with $\tau=3/2$
\cite{ziff}. 

Now, for given $d$ and $D$, and plugging the expected small $s$ behavior $f(s)
\sim s^{-\tau}$ into Eq. (\ref{scaleq}), one first gets that
$\tau<1+\lambda=1+d/D$. Then,  matching the behavior of both sides of Eq.
(\ref{scaleq})  \cite{vandong2,nous}, one finds,
\begin{equation} \label{tau}
\tau=2-\int_0^{\infty} f(x) x^\frac{d}{D} dx
\end{equation}
If $\alpha>\tau-1$ we obtain by multiplying Eq. (\ref{scaleq}) by $x^\alpha$
and integrating \cite{vandong2,vandong3,nous}:
\begin{eqnarray}\label{eqal}
&2(1-\alpha) \int x^\alpha f(x) \,dx=\nonumber\\
&\int f(x)f(y)(x^\frac{1}{D}+y^\frac{1}{D})^d 
\left[ x^\alpha+y^\alpha -\right.\left.(x+y)^\alpha\right]\,dx dy
\end{eqnarray}  

By studying the large s behavior of Eq. (\ref{scaleq}), one can show  that
$f(s)$ decays as $c_\infty s^{-\frac{d}{D}} e^{-s}$ with $c_\infty ^{-1}=
\int_0^{1/2}(x^\frac{1}{D}+(1-x)^\frac{1}{D})^d
x^{-\frac{d}{D}}(1-x)^{-\frac{d}{D}}dx$.

{\em Exact bounds and estimates -} We first show that $\tau\geq 1$ if
$d\geq1$. Suppose $\tau<1$ and consider Eq. (\ref{eqal}) with $\alpha=0$. For
$d\geq 1$, we have $(x^\frac{1}{D}+y^\frac{1}{D})^d\geq
x^\frac{d}{D}+y^\frac{d}{D}$, which leads to  $\int f(x)dx \leq \int f(x)dx
\int f(x)x^\frac{d}{D}dx$, hence $1\geq 2-\tau$, which is contradictory.
Notice that Eq. (\ref{eqal}) with $\alpha=2$ for $d=1$ and $D=1$ leads to
$\int x^2 f(x)dx= 2 (\int x^2f(x)dx)(\int xf(x)dx)$, and we recover the exact
result $\tau=2-\int x f(x)\,dx=3/2$ \cite{ziff} in a very simple way.    

We now introduce an extremely simple method of getting lower and upper bounds
for $\tau$. We rely on Eq. (\ref{eqal}) valid for $\alpha>\tau-1$. Combining
Eq. (\ref{tau}) and  (\ref{eqal}), we get:
\begin{equation} \label{mean}
\tau=2-(1-\alpha)\frac{\int g(x,y)\,dxdy}{\int g(x,y)A(x/y)\,dxdy}
\end{equation} 
where $A(u)=(1+u^\alpha-(1+u)^\alpha) (1+u^\frac{1}{D})^d /(u^\alpha+ u^{\frac
d D})$ satisfies $A(u)=A(1/u)$ and $g(x,y)=(x^\alpha y^{\frac d D}+x^{\frac d
D}y^\alpha)f(x)f(y)$. The ratio in Eq. (\ref{mean}) can then be interpreted as
the inverse of a kind of $average$ of $A(x/y)$ with the weight $g(x,y)$.  For
a given $\alpha\leq d/D$, we determine numerically the lower and upper bounds
$m_\alpha$ and $M_\alpha$ of the function $A(u)$.  Using Eq. (\ref{mean}),
this gives $2-(1-\alpha)/m_\alpha\leq \tau\leq 2-(1-\alpha)/M_\alpha$. We then
choose the best values of $\alpha\leq d/D$ compatible with $\alpha>\tau-1$
leading to the tightest bounds. This allows us to greatly improve the exact
bounds given in \cite{vandong2,vandong3} for $d=1$ and to obtain new such
bounds for $d>1$. For instance for the physically interesting cases (see
below) $(d=1,D=2)$, $(d=1,D=4)$ and $(d=2,D=4)$ we respectively found
$1.084\leq\tau\leq 1.147 $, $1\leq\tau\leq 1.075$ (compared to $1\leq\tau\leq
1.28 $ and $1\leq\tau\leq 1.109 $ in \cite{vandong3}) and $1.25\leq\tau\leq
1.5$.

It is also possible to obtain good estimates by evaluating the `average' in
Eq. (\ref{mean}) using a reasonable trial weight function $g(x,y)$ instead of
the unknown exact one. A parameter free choice is obtained by replacing in the
above expression of $g(x,y)$ the exact $f(x)$ by $x^{-\tau}\exp(-x)$ which has
the correct leading asymptotic for small $x$ (by definition of $\tau$) and the
expected exponential large $x$ decay \cite{vandong1,nous}. This form is known
to be a good approximation of the actual $f(x)$ obtained in simulations
\cite{krivitsky}, and is even obtained in exactly solvable models \cite{ziff}.
The simplest  method is to determine $\tau$ self-consistently from Eq.
(\ref{mean}), for instance  with $\alpha=d/D$, but the  result depends on the
chosen $\alpha$ and may even violate exact bounds. A much better and hardly
more intricate method is to choose a sample of values of $\alpha$, and
minimize an error function measuring the violation of the corresponding Eq.
(\ref{mean}) \cite{nous}. This method can be systematically improved by
allowing for $n$ free `fitting' parameters (including $\tau$ itself) in the
trial weight $g(x,y)$. Using the simplest $\chi^2$ form for the error function
with a trial function $f(x)=(x^{-\tau}+cx^{-\frac{d}{D}} )e^{-x}$ (to take
into account the exact decay at large $x$), we obtain \cite{nous}:
$\tau\approx 1.111$, $\tau\approx 1.016$ and $\tau\approx 1.431$ in the three
cases considered above.  The case $d=1$, $D>1$ has been numerically
investigated by means of time series in \cite{song}. The authors of
\cite{song} found  $\tau\approx 1.11\sim 1.12$ for $D=2$ (using data in the
text of \cite{song}) and to $\tau\approx 1.06$ for $D=4$ (as roughly extracted
from  Fig. 3 in \cite{song}), in fair agreement with our bounds and estimates.
We shall find later that our estimate for $d=2$ and $d=4$ is in good agreement
with a non perturbative calculation for $\tau$ and with a perturbative
estimate. Our {\em variational} method requires very few CPU time and is
straightforwardly implemented compared  to methods used in
\cite{krivitsky,song}.

{\em Perturbative expansions for $d< 1$ -} Now we use the exactly solvable
limits $d=0$ and $D=\infty$ as a basis for a perturbative expansion.

First, we consider the limit $d \rightarrow 0$. We expand $f$ in series of
$d$: $f(x)=f_0(x)+d f_1(x) + O(d^2)$. A systematic way of expanding  $\tau$
would be to write down a linear (self-consistent) differential equation for
$f_1$ to solve it and plug the result into (\ref{tau}). This method is used in
\cite{nous} to compute the next order $O(d^2)$. However, as far as the first
order is concerned we can get it without solving for $f_1$. By developing the
integral expression of $\tau$, Eq. (\ref{tau}), we get: $\tau= 2-\int
f(x)x^\frac{d}{D} dx= -d/D \int f_0(x)\ln x \,dx - d\int f_1 +O(d^2)$. Now we
develop both sides of Eq. (\ref{eqal}) with $\alpha=0$ to get an equation for
$\int f_1$ : $\int f_1= \frac{1}{2}\int f_0(x)f_0(y)
\ln(x^\frac{1}{D}+y^\frac{1}{D})dx dy- (\int f_0) (\int f_1)$, hence $\int
f_1=-\int e^{-x-y} \ln(x^\frac{1}{D}+y^\frac{1}{D}) dxdy$. After a simple
calculation we get:
\begin{equation}\label{petitd}
\tau= 2d \int_0^1 \ln\left(1+\left(\frac{1-u}{u}\right)^\frac{1}{D}\right)\, 
du +O(d^2).
\end{equation}
Let us mention that we can generalize this result to any homogeneous kernel of
the form: $(g(x,y))^d$, leading to, $\tau= 2d\int_0^1 \ln g(1,\frac{1-u}{u})
du +O(d^2)$.

For $D=1$, we get $\tau=2d+O(d^2)$, in good agreement with direct numerical
integration of Smoluchowski's equation performed by Krivitsky
\cite{krivitsky}, who obtained $\tau\approx 0.2$ for $d=0.1$ and $\tau\approx
0.38$ for $d=0.2$. This result for $D=1$ also coincides up to order $O(d)$
with the best inequalities for $\tau$ that we obtained above (as already
observed in \cite{vandong3}), but not for other values of $D$ \cite{nous}. 

The order $O(d^2)$ requires the computation of $f_1$ which satisfies a
solvable linear second order equation. This cumbersome calculation will be
presented in \cite{nous}. However, in the special case $D=1$  it is possible
to obtain explicitly the $O(d^2)$ term by expanding Eq. (\ref{eqal}) for
$\alpha=d/D$.  We obtain, $\tau=2d+(\frac{\pi^2}{3}-4)d^2 +O(d^3)$
\cite{nous}. For $d=0.2$, we get  $\tau \approx 0.372$ compared to the already
mentioned $\tau \approx 0.38$, whereas for $d=0.4$, we get $\tau \approx
0.686$ compared to $\tau\approx 0.7$ found numerically in \cite{krivitsky}.

Now, we perform an expansion in powers of $1/D$ for $d\leq 1$, expanding
$f(x)=f_{\infty}(x)+\frac{1}{D} f_1(x) + \frac{1}{D^2}f_2(x)+O(1/D^3)$.  We
use exactly the same method: we develop Eq. (\ref{eqal}) with $\alpha=0$ in
powers of $1/D$, and plug the result in Eq. (\ref{tau}), yielding a vanishing
first order term and a nontrivial second order term:
\begin{equation} \label{alD2}
\tau =2-2^{1-d} + \frac{\pi^2 2^{-d} d(1-d)}{12D^2} +O(\frac{1}{D^3})
\end{equation}
Once again we were able to obtain a highly nontrivial expansion of $\tau$
without solving for $f_1$ and $f_2$ themselves, although this can also be
achieved this way \cite{nous}. Note that in the limit of large $D$ $and$ small
$d$, Eq. (\ref{petitd}) and (\ref{alD2}) coincide up to order $O(d/D^2)$.

{\em Perturbative estimate for $d>1$ -} In the case $d\geq 1$, we have shown
that $\tau\geq 1$ and since $\tau<1+d/D$, we see that $\tau\to 1$ for
$D\to\infty$ and finite $d>1$. The previous perturbation is not valid because
$f_1$ is non integrable. Nevertheless we can try to obtain an estimate of
$\tau$ in the following way: we make the ansatz $f\sim
f_\infty+c/s^{1+\varepsilon} e^{-s}$ when $s\rightarrow 0$. We plug it into
Eq. (\ref{tau}) and Eq. (\ref{eqal}) for $\alpha =d/D$, and after some algebra
\cite{nous} we see  that for consistency $\varepsilon$ must be of order $1/D$
and that $c=(1-2^{1-d})(d/D-\varepsilon)$, and eventually that
$\varepsilon=\kappa/D +O(1/D^2)$ where $\kappa$ is the solution of the
nonlinear equation:
\begin{equation}\label{estimate}
\frac{2}{1+2^{1-d}} = \int_0^1 (1+u^\frac{1}{d-\kappa})^d dv
\end{equation}

This equation always has a solution consistent with the exact bound
$1<\tau<1+d/D$. For instance in the case $d=2$, $D=4$ we obtain $\tau \approx
1.462$, which is highly consistent with the previous  estimate  $\tau\approx
1.431$. Though it is still of order $1/D$, the obtained perturbative estimate
depends on the choice of $\alpha$. $\alpha=d/D$ seems however to be the most
natural choice. 

In $d=1$, $c$ vanishes and we do not learn much. Notice that we   have shown
that all terms of the $d<1$ series for $\tau$ in powers of $1/D$ vanish for
$d\rightarrow1$, as can be seen in Eq. (\ref{alD2}) for the two leading ones.
Thus, the correction to  $\tau=1$ for large $D$ may be {\it non perturbative}
in $d=1$, which would again rule out the estimate $\tau=1+1/2D$ of
\cite{Leyvraz}, which also  violates our rigorous inequalities
\cite{vandong3,nous}.

If we now take the $d\to\infty$ limit in Eq. (\ref{estimate}), we obtain,
$\tau\simeq 1+\lambda -2^{-d} \lambda$ ($\lambda =d/D$), a non perturbative
behavior  in $d$ which is to be related to the  results below. 

{\em Large d and D -} We now present a non perturbative calculation in the
limit of large $d$ $and$ $D$, keeping the ratio $\lambda=d/D$ fixed. In this
limit, the kernel can be written,
\begin{equation} \label{kerpro}
(x^\frac{1}{D} + y^\frac{1}{D})^d=2^d(xy)^{\frac{\lambda}{2}}(1+O(d/D^2))
\end{equation}
and surprisingly transforms into the celebrated `product' kernel
\cite{meakinrev,vandong2,krivitsky,song,Leyvraz}. Assuming scaling (a still
controversial subject \cite{krivitsky,nous}), one can easily show that
$\tau=1+\lambda=1+d/D$ \cite{vandong2} (see also Eq. (\ref{genscal1}) and
(\ref{genscal2}) and the discussion below them, as it corresponds to
$\mu=\lambda/2>0$). We have shown that including higher order corrections in
power of $1/D$ does not change the value of $\tau$ such that the correction to
$\tau=1+\lambda$ is certainly non perturbative. In fact, we can estimate this
correction by assuming that for finite $d$ and $D$, $f(s)\sim
c_\lambda/s^{1+\lambda -\varepsilon_d}$ for $s\to 0$. Plugging this estimate
in Eq. (\ref{tau}) with the limit kernel of Eq. (\ref{kerpro}), we first get
$\varepsilon_d\approx 2^{-d}c_\lambda/(1-\lambda)$. $c_\lambda$ can be
determined by matching the coefficients of the leading terms in Eq.
(\ref{scaleq}) using the kernel of Eq. (\ref{kerpro}). After a straightforward
calculation, one gets $c_\lambda$ in the $d\to\infty$ limit: 
\begin{eqnarray}
\tau=1+\lambda-2^{1-d}I_\lambda^{-1}\label{c},\quad
c_\lambda={2(1-\lambda)}{I_\lambda}^{-1}\\
I_\lambda=\int_0^1 
[u(1-u)]^{-1-\lambda/2}\left[u^\lambda+(1-u)^\lambda-1\right]\,du
\nonumber
\end{eqnarray}
We thus find a non perturbative (exponentially small) correction to $\tau$ in
the large $d$ and large $D$ limit, consistent with the result obtained above
for $d>1$ and large $D$. Note that Eq. (\ref{c}) is also consistent with the
exact result that $\tau\to 1$ as $D\to\infty$ for finite $d>1$, a result that
we obtain by setting $\lambda=0$ (as $I_\lambda$ diverges). In the case $d=2$
and $D=4$ of interest below, we already gave the estimate $\tau\approx 1.431$,
whereas Eq. (\ref{c}) leads to $\tau\approx 1.428$. 

Physical applications of these results to massive particle aggregation systems
and the generalization to other physically relevant kernels (as the one
applying to the systems described in \cite{meakindrop,familymeakin}) will be
presented elsewhere \cite{nous}. In this Letter, we would like to present an
original application outside of this field of massive particle aggregation,
namely the dynamics of vortices in two-dimensional decaying turbulence.

Recently, a statistical numerical model has been introduced
\cite{McWilliams,Benzi} which describes the dynamics and the merging of
vortices with the assumption that the typical core vorticity $\omega$ and the
total energy $E\sim \int v^2\,d^2x\sim \sum_i\omega^2 R_i^4$ are conserved
($R_i$ is the radius of the $i$-th vortex) throughout the merging processes.
This model reproduces the main features observed in direct numerical
simulations (see \cite{McWilliams,Benzi} for details). For instance, after
noting that a distribution of vortex radii satisfying $P(R)\sim R^{-\beta}$ is
equivalent to a Gaussian energy spectrum $E(k)\sim k^{\beta-6}$ \cite{Benzi},
the simulation of this model was able to reproduce the fact that starting from
a Batchelor spectrum $E(k)\sim k^{-3}$ ($\beta=3$), the system evolves
systematically to a steeper spectrum $E(k)\sim k^{-\gamma}$ with
$\gamma=6-\beta$ in the range $\gamma\approx 3\sim 5$ \cite{Benzi}.

Now, one expects that the collision kernel between two vortices is somewhat
intermediate between the ballistic hard-disk form $\sigma\sim (R_1+R_2)$
\cite{song}, and the totally uncorrelated form $\sigma\sim (R_1+R_2)^2$ (where
the probability of colliding is proportional to the probability that two
randomly placed vortices overlap, see also below Eq. (\ref{conser}))
\cite{nous}. Thus, one can describe approximately the decay of vortices due to
mergings by means of Eq. (\ref{scaleq}) with $1\leq d \leq 2$ and $D=4$, as
two colliding vortices merge into a new one with $R=(R_1^{4}+R_2^{4})^{1/4}$
in order to conserve energy and core vorticity. One thus expects a power law
radius distribution $P(R)\sim R^{-\beta}$, with $\beta=D(\tau-1)+1$ and $\tau$
given by our model. We find values of $\gamma$ ranging from $\gamma\approx
3.3$ for $d=2$ (taking $\tau\approx 1.43$) to $\gamma\approx 4.94$ (taking
$\tau\approx 1.016$) for $d=1$, in good qualitative agreement with observed
exponents. As also found in direct simulations, the actual exponent (and here
the value of the effective correct $d$) could depend on the actual initial
conditions ($\omega$, area occupied by the vortices $\sim$ enstrophy). Note
that the limit Batchelor case $\gamma=3$, is obtained when taking the naive
strict upper bound $\tau=1+d/D$ with $d=2$ and $D=4$.

In conclusion, we have introduced a systematic scheme to obtain exact bounds
and good estimates for the polydispersity exponent in aggregation models. We
have also implemented perturbative and non perturbative expansions found to be
in good agreement with already known numerical results when available.
Finally, this kind of calculations generalizes to other interesting kernels,
with possible physical applications in the field of droplet deposition
\cite{familymeakin,nous} or even two-dimensional decaying turbulence as
briefly mentioned in the present Letter.

We are very grateful to F. Leyvraz and S. Redner for valuable comments on  the
manuscript.

\end{multicols}
\end{document}